\documentclass[submission,copyright,creativecommons]{eptcs}
\providecommand{\event}{PxTP~2015} 
\usepackage{breakurl}             

\usepackage{fixltx2e}
\usepackage{graphicx}
\usepackage{longtable}
\usepackage{float}
\usepackage{wrapfig}
\usepackage{soul}
\usepackage{textcomp}
\usepackage{marvosym}
\usepackage{wasysym}
\usepackage{latexsym}
\usepackage{amssymb}
\usepackage{verbatim}
\usepackage[T1]{fontenc}
\usepackage{tikz}
\usetikzlibrary{shapes,trees,positioning,arrows}
\hypersetup{colorlinks=true}

\newcommand\const[1]{\ensuremath{\mathsf{#1}}}
\newcommand\Type{\const{Type}}

\newcommand\type{\const{type}}
\newcommand\prop{\const{prop}}
\newcommand\bool{\const{bool}}
\newcommand\nat{\const{nat}}
\newcommand\arrow{\const{arrow}}
\newcommand\term{\const{term}}

\newcommand\proof{\const{proof}}

\newcommand\rewritesto{\leadsto}

\newcommand\lpm{$\lambda\Pi$-calculus modulo rewriting}
\newcommand\tlpm{the \lpm{}}

\title{Mixing HOL and Coq in Dedukti \\ (Extended Abstract)}

\def\titlerunning{Mixing HOL and Coq}

\author{Ali Assaf\institute{INRIA Paris-Rocquencourt, France}\institute{\'Ecole Polytechnique, France}
  \and Rapha\"el Cauderlier\institute{INRIA Paris-Rocquencourt, France}\institute{Laboratoire CEDRIC, CNAM, France}}

\def\authorrunning{A. Assaf and R. Cauderlier}

\begin{document}

\maketitle

\begin{abstract} We use Dedukti as a logical framework for interoperability. We
use automated tools to translate different developments made in HOL and in Coq
to Dedukti, and we combine them to prove new results. We illustrate our
approach with a concrete example where we instantiate a sorting algorithm
written in Coq with the natural numbers of HOL.


\end{abstract}

\section{Introduction}\label{sec-introduction}

Interoperability is an emerging problem in the world of proof systems.
Interactive theorem provers are developed independently
and cannot usually be used together effectively. The theorems of one system can
rarely be used in another, and it can be very expensive to redo the proofs
manually.
Obstacles for a large-scale interoperability are many, ranging from differences in the logical theory and the representation of data types,
to the lack of a standard and effective way of retrieving proofs.
For systems based on a common logical formalism, exchange formats for proofs have appeared
like the TPTP derivation format \cite{TPTPderivation} for traces of automated first-order
theorem provers and OpenTheory \cite{hurd_opentheory_2011} for HOL interactive theorem provers.
However, combining systems working in different logical theories is harder.

A solution to this problem is to use a logical framework. The idea is to have a small and simple
language that is expressive and flexible enough to define various logics and to
faithfully express proofs in those logics, at a relatively low cost.
Translating all the different systems to this common framework is a first step
in bringing them closer together. This is the idea behind LF \cite{harper_framework_1993}, implemented in
Twelf \cite{pfenning_system_1999}, which has been used as a framework for interoperability in various projects \cite{schurmann_executable_2006,horozalrabe2011}.

We propose to use a variant of Twelf called Dedukti. The reason for using Dedukti is that it implements an extension of LF called the \emph{\lpm{}} \cite{boespflug_lambda-pi-calculus_2012, cousineau_embedding_2007}, which adds term rewriting to the calculus. This extension not only allows for a more compact representation of proofs, but also enables the encoding of richer theories, such as the calculus of constructions. This cannot be done in LF efficiently because computation would have to be represented as a relation and every conversion made explicit. We thus use Dedukti as our logical framework.

Several tools have been developed to translate the proofs of
various systems to Dedukti  \cite{boespflug_coqine_2012,assaf_translating_2014,delahaye_zenon_2013,burel_shallow_2013}. The translations are based on the encodings of Cousineau and Dowek in \tlpm{} \cite{cousineau_embedding_2007}. The proofs, represented as terms of \tlpm{}, can be checked independently by Dedukti, adding another layer
of confidence over the original systems. This approach has been successfully
used to verify the formalization of several libraries and the proof traces of
theorem provers on large problem sets (of the order of several gigabytes).

In this paper, we take one step further and show that we can combine the proofs
coming from different systems in this same framework. A theorem can therefore be
split into smaller blocks to be proved separately using
different systems, and large libraries formalized in one system can be reused
for the benefit of developments made in another one.

This approach has several advantages. First, we can use Dedukti as an independent proof checker. The \lpm{} is fairly simple, and the kernel implementation is relatively small \cite{saillard_dedukti_2013,saillard_towards_2013} compared to systems like Coq. The soundness and completeness of the translations have been studied and proved \cite{assaf_conservativity_2015,cousineau_embedding_2007,dowek_models_2014}, giving us further confidence. Compared to direct one-to-one translations \cite{kaliszyk_scalable_2013,keller_importing_2010,obua_importing_2006,naumov_hol/nuprl_2001}, we avoid the quadratic blowup of the number of translations needed to translate $n$ systems. In that scenario, if a new proof system enters the market, we would need to design $n$ new translations. Moreover, some systems such as Coq have complex foundations that are difficult to translate to other formalisms. Another possibility would be to compose existing translations, provided that they are scalable and composable. This avenue has not been investigated. In our approach, we instead translate the different systems to one common framework. We do not propose translations back into other systems, as we can use Dedukti as a low-level assembly language, akin to machine language when we compile and link programs coming from different programming languages.

\subsection*{Contributions}

We used Holide and Coqine to translate proofs of HOL \cite{harrison_hol_2009} and Coq \cite{coq_development_team_coq_2012}, respectively, to
Dedukti. We examined the logical theories behind those two systems to determine
how we can combine them in a single unified theory while addressing the problems
mentioned above. Finally, we used the resulting theory to certify the
correctness of a sorting algorithm involving Coq lists of HOL natural
numbers. Our code is available online at \url{http://dedukti-interop.gforge.inria.fr/}.

\begin{center}
\bigskip
\includegraphics[]{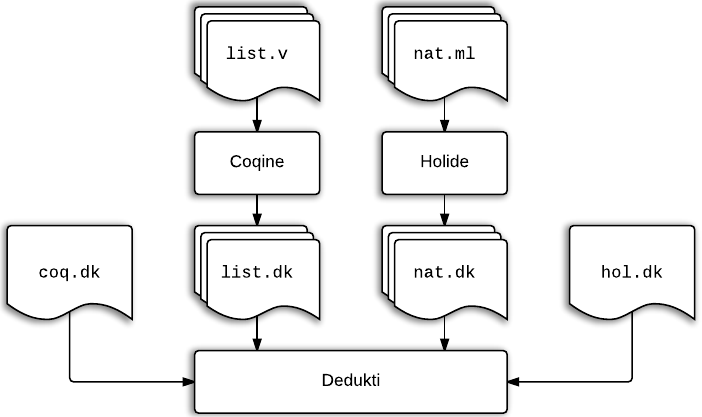}
\end{center}

\section{Tools used}
\label{sec-tools}

\subsection*{Dedukti}

Dedukti\footnote{Available at: \url{http://dedukti.gforge.inria.fr/}} is a
functional language with dependent types based on \tlpm{} \cite{saillard_dedukti_2013,saillard_towards_2013}. The
type-checker/interpreter for Dedukti is called \texttt{dkcheck}. It accepts files
written in the Dedukti format (\texttt{.dk}) containing declarations,
definitions, and rewrite rules, and checks whether they are well-typed.

Following the LF tradition, Dedukti acts as a logical framework to define logics
and express proofs in those logics. The approach consists in representing
propositions as types and proofs as terms inhabiting those types, as in the
Curry-Howard correspondence. Assuming the representation is correct, a proof is
valid if and only if its corresponding proof term is well-typed. That way we can
use Dedukti as an independent proof checker.

\subsection*{Holide}

Holide\footnote{Available at: \url{https://www.rocq.inria.fr/deducteam/Holide/}}
translates HOL proofs to the Dedukti language. It accepts proofs in the
OpenTheory format (\texttt{.art}) \cite{hurd_opentheory_2011}, and generates
files in the Dedukti format (\texttt{.dk}). These files can then be verified by
Dedukti to check that the proofs are indeed valid. The translation is described in
detail in \cite{assaf_translating_2014}.

The generated files depend on a handwritten file called \texttt{hol.dk}. This
file describes the theory of HOL, that is the types, the terms, and the
derivation rules of HOL. The types of HOL are those of the simply-typed
$\lambda$-calculus. We represent them as terms of type $\type$ (not to be
confused with $\Type$, the ``type of types'' of Dedukti). We represent the
propositions as terms of type $\bool$.
\[
\begin{array}{llllll}
\type & : & \Type. \qquad & \term & : & \type \rightarrow \Type. \\
\bool & : & \type. \qquad & \proof & : & \term\ \bool \rightarrow \type. \\
\arrow & : & \type \rightarrow \type \rightarrow \type. \qquad & ... \\
\end{array}
\]

\subsection*{Coqine}

Coqine\footnote{Available at:
\url{http://www.ensiie.fr/~guillaume.burel/blackandwhite_coqInE.html.en}} translates
Coq proofs to the Dedukti language. It takes the form of a Coq plugin that can
be called to export loaded libraries (\texttt{.vo}) to generate files in the
Dedukti format (\texttt{.dk}). These files can then be verified by Dedukti to
check that the proofs are indeed valid.

A previous version of the translation is described in
\cite{boespflug_coqine_2012}. However, that translation is outdated, as it does
not support the universe hierarchy and universe subtyping of Coq. A \emph{universe} is just another name for a ``'type of types''. To avoid paradoxes, they are stratified into an infinite hierarchy \cite{barendregt_lambda_1992}, but that hierarchy is ignored by the first implementation of Coqine. The
translation has since been updated to support both features following the ideas
in \cite{assaf_calculus_2014}, although some other features such as the module system are still missing.

The generated files depend on a handwritten file describing the theory of the
calculus of inductive constructions (CIC) called \texttt{coq.dk}. There is a
type $\prop$ that represents the universe of propositions and a type $\type\ i$
for every natural number $i$ that represents the $i$th universe of types. We
will write $\const{type}_i$ and $\const{term}_i$ for, respectively, $\const{type}\ i$ and
$\const{term}\ i$.
\[
\begin{array}{llllll}
\type & : & \nat \rightarrow \Type. \qquad & \const{term} & : & \Pi i : \nat.\ \type\ i \rightarrow \Type. \\
\prop & : & \Type. \qquad & \const{proof} & : & \prop \rightarrow \Type. \\
... \\
\end{array}
\]

\section{Mixing HOL and Coq}
\label{sec-mixing}

HOL and Coq use very different logical theories. The first is based on Church's
simple type theory, is implemented using the LCF approach, and its proofs are
built by combining sequents in a bottom-up fashion. The second is based on the
calculus of inductive constructions and checks proofs represented as $\lambda$-terms
in a top-down fashion. Translating these two systems to Dedukti was a
first step to bringing 
them closer together, but there are still important
differences that set them apart. In this section, we examine these differences
and show how we were able to bridge these gaps.

\subsection*{Type inhabitation}

The notion of types is different between HOL and Coq. In HOL, types are those of
the simply-typed $\lambda$-calculus where every type is inhabited. In
contrast, Coq allows the definition of empty types, which in fact play an
important role as they are used to represent falsehood. A na\"ive reunion of the
two theories would therefore be inconsistent: the formula $\exists x : \alpha,
\top$, where $\alpha$ is a free type variable, is provable in HOL but its
negation $\neg \forall \alpha : \const{Type}, \exists x : \alpha, \top$ is
provable in Coq.

Instead, we match the notion of HOL types with that of Coq's \emph{inhabited}
types, as done by Keller and Werner \cite{keller_importing_2010}. We define inhabited types in
the Coq module \texttt{holtypes}:
\begin{verbatim}
    Inductive type : Type := inhabited : forall (A : Type), A -> type.
\end{verbatim}
It is then easy to prove in Coq that given inhabited types $A$ and $B$,
the arrow type $A \to B$ is also inhabited:
\begin{verbatim}
    Definition carrier (A : type) : Type :=
      match A with inhabited B b => B end.
    Definition witness (A : type) : carrier A :=
      match A with inhabited B b => b end.
    Definition arrow (A : type) (B : type) : type :=
      inhabited (carrier A -> carrier B) (fun _ => witness B).
\end{verbatim}
This is all that we need to interpret \const{hol.type}, \const{hol.term}, and
\const{hol.arrow} using rewrite rules:
\[
\begin{array}{lll}
\const{hol.type} & \rewritesto & \const{coq.term}_1\ \const{holtypes.type}. \\
\const{hol.arrow}\ a\ b\ & \rewritesto & \const{holtypes.arrow}\ a\ b. \\
\const{hol.term}\ a & \rewritesto & \const{coq.term}_1\ (\const{holtypes.carrier}\ a). \\
\end{array}
\]

\subsection*{Booleans and propositions}

In Coq, there is a clear distinction between booleans and propositions. Booleans
are defined as an inductive type \const{bool} with two constructors \const{true}
and \const{false}. The type \const{bool} lives in the universe \const{Set} (which is another name for the universe $\Type_0$). In
contrast, following the Curry-Howard correspondence, propositions are
represented as types with proofs as their inhabitants. These types live in the
universe \const{Prop}.
Both \const{Set} and \const{Prop} live in the universe $\const{Type}_1$. As a
consequence, \const{Prop} is not on the same level as other types such as
\const{bool} or \const{nat} (the type of natural numbers), a notorious feature
of the calculus of constructions. Moreover, since Coq is an intuitionistic
system, there is no bijection between booleans and propositions. The excluded
middle does not hold, though it can be assumed as an axiom.

In HOL, there is no distinction between booleans and propositions and they are
both represented as a single type \const{bool}. Because the system is
classical, it can be proved that there are only two inhabitants $\top$ and
$\bot$, hence the name. Moreover, the type \const{bool} is just another simple
type and lives on the same level as other types such as \const{nat}.

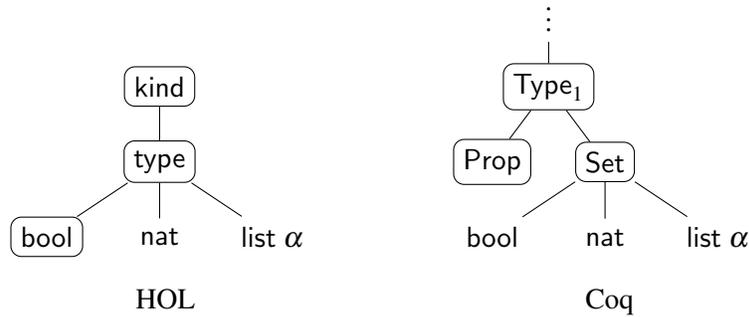
\begin{figure}
\begin{center}
{
  \setlength{\tabcolsep}{25pt}
  \vspace{-25pt}
  \begin{tabular}{cc}

  \begin{tikzpicture}
  \tikzstyle{universe}=[draw,rounded corners, level distance=10cm]
  \tikzstyle{level 1}=[level distance=10mm]
  \draw node [universe] {$\const{kind}$}
    child {node [universe] {$\const{type}$}
      child {node [universe] {$\const{bool}$}}
      child {node {$\const{nat}$}}
      child {node {$\const{list}\ \alpha$}}};
  \end{tikzpicture}
  \smallskip

  &

  \begin{tikzpicture}
  \tikzstyle{universe}=[draw,rounded corners]
  \tikzstyle{level 1}=[level distance=10mm]
  \draw node {$\vdots$}
      child {node [universe] {$\const{Type}_1$}
        child {node [universe] {$\const{Prop}$}}
        child {node [universe] {$\const{Set}$}
          child {node {$\const{bool}$}}
          child {node {$\const{nat}$}}
          child {node {$\const{list}\ \alpha$}}}};
  \end{tikzpicture}
  \smallskip

  \\

  HOL & Coq

  \end{tabular}
  \vspace{-15pt}
}
\end{center}
\caption{
  Booleans and propositions in HOL and Coq. Boxes represent universes.}
\label{fig:bool-vs-prop}
\end{figure}

To combine the two theories, one must therefore reconcile the two pictures in
Figure \ref{fig:bool-vs-prop}, which show how the types of HOL and Coq are organized.\footnote{Since $\bool$ is the type of propositions, and propositions are the types of proofs in the Curry-Howard correspondence, $\bool$ can be viewed as a universe \cite{barendregt_lambda_1992,geuvers_logics_1993}.} One solution is to interpret the types of HOL as
types in $\const{Set}$. To do this, we must rely on a reflection mechanism that
interprets booleans as propositions, so that we can retrieve the theorems of HOL
and interpret them as theorems in Coq. In our case, it consists of a function
\const{istrue} of type $\const{hol.bool} \rightarrow \const{coq.prop}$, which we
use to define $\const{hol.proof}$:
\[
\const{hol.proof}\ b \rewritesto \const{coq.proof}\ (\const{istrue}\ b).
\]

Another solution is to translate \const{hol.bool} as \const{coq.prop}. To do
this, we must therefore translate the types of HOL as types in $\const{Type}_1$
instead of $\const{Type}_0$. In particular, if we want to identify
$\const{hol.nat}$ and $\const{coq.nat}$, we must have $\const{coq.nat}$ in
$\const{Type}_1$. Fortunately, we have this for free with cumulativity since any
element of $\const{Type}_0$ is also an element of $\const{Type}_1$.

We choose the first approach as it is more flexible and places less restrictions
(e.g.~regarding $\const{Prop}$ elimination in Coq) on what we can do with
booleans. In particular, it allows us to build lists by case analysis on
booleans, which is needed in our case study.

\section{Case study: sorting Coq lists of HOL numbers}
\label{sec-example}

We proved in Coq the correctness of the insertion sort algorithm on polymorphic
lists and we instantiated it with the canonical order of natural numbers
defined in HOL. More precisely, on the Coq side, we defined polymorphic lists,
the insertion sort function, the \const{sorted} predicate, and the
\const{permutation} relation. We then proved the following two theorems:
\begin{verbatim}
    Theorem sorted_insertion_sort: forall l, sorted (insertion_sort l).
    Theorem perm_insertion_sort: forall l, permutation l (insertion_sort l).
\end{verbatim}
with respect to a given (partial) order:
\begin{verbatim}
    Variable A : Set.
    Variable compare : A -> A -> bool.
    Variable leq : A -> A -> Prop.
    Hypothesis leq_trans : forall a b c, leq a b -> leq b c -> leq a c.
    Hypothesis leq_total : forall a b, if compare a b then leq a b else leq b a.
\end{verbatim}
The order comes in two flavors: a relation \const{leq} used for proofs,
and a decidable version \const{compare} which we can destruct for building
lists. The totality assumptions relates \const{leq} and \const{compare} and
can be seen as a specification of \const{compare}.

On the HOL side, we used booleans, natural numbers and the order relation on
natural number as defined in the OpenTheory packages \texttt{bool.art} and \texttt
{natural.art}. By composing the results, we obtain two Dedukti theorems:
\[
\begin{array}{l}
\Pi l : \const{coq.term}_1\ (\const{coq\_list}\ \const{hol\_nat}).\ \const{proof}\ (\const{sorted}\ (\const{insertion\_sort}\ \const{compare}\ l)). \\
\Pi l : \const{coq.term}_1\ (\const{coq\_list}\ \const{hol\_nat}).\ \const{proof}\ (\const{permutation}\ l\ (\const{insertion\_sort}\ \const{compare}\ l)). \\
\end{array}
\]

The composition takes place in a Dedukti file named \texttt{interop.dk}. This file takes care of matching the interfaces of the proofs coming from Coq with the proofs coming from HOL. Most of the work went into proving that HOL's comparison is indeed a total order in Coq:
\[\Pi m~n : \const{holtypes.carrier}~\const{hol\_nat}.\ \const{if}~(\const{compare}~m~n)~\const{then}~m \leq n~\const{else}~n \leq m.\]
We prove it using the following theorems from OpenTheory:
\[
\begin{array}{ll}
\forall m~n : \const{hol\_nat}.\ m < n \Rightarrow m \leq n \\
\forall m~n : \const{hol\_nat}.\ m \not\leq n \Leftrightarrow n < m \\
\end{array}
\]
and some additional lemmas on $\const{if} \ldots \const{then} \ldots \const{else}$.
Because of the verbosity of Dedukti and small style differences between HOL
and Coq, this proof is long (several hundreds of lines) for such a simple fact.
However, most of it is first-order reasoning and we believe that it could be automatically proved by the theorem prover Zenon~\cite{zenon} which can output proofs in the Dedukti format~\cite{Zenon-Modulo,delahaye_zenon_2013}.

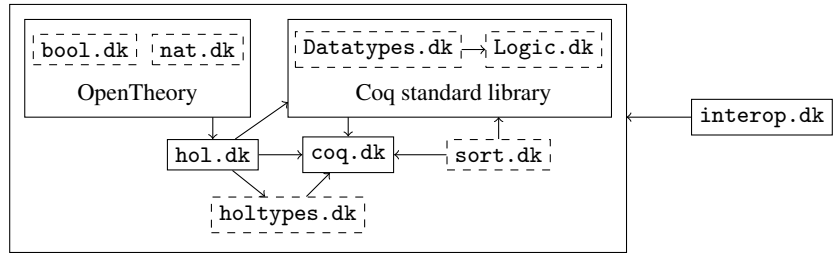
\begin{figure}
\begin{center}
\begin{tikzpicture}
\footnotesize
\tikzstyle{manual}=[draw]
\tikzstyle{auto}=[draw,style=dashed]
\tikzstyle{dep}=[->]

\node [manual] (coqdk) at (0, 0) {\texttt{coq.dk}};

\node [auto] (logicdk) at (2.6, 1.4) {\texttt{Logic.dk}};
\node [auto] (datatypesdk) at (0.4, 1.4) {\texttt{Datatypes.dk}} edge [dep] (logicdk);

\draw [dep] (0, 0.5) -- (coqdk);

\node [auto] (holtypesdk) at (-0.8, -0.8) {\texttt{holtypes.dk}} edge [dep] (coqdk);
\node [auto] (sortdk) at (2, 0) {\texttt{sort.dk}} edge [dep] (coqdk) edge [dep] (2, 0.5);

\node [manual] (holdk) at (-1.8, 0) {\texttt{hol.dk}}
   edge [dep] (-0.8, 0.7)
   edge [dep] (coqdk)
   edge [dep] (holtypesdk);

\node [auto] (booldk) at (-3.5, 1.4) {\texttt{bool.dk}};
\node [auto] (natdk) at (-2, 1.4) {\texttt{nat.dk}};
\draw [dep] (-1.8, 0.5) -- (holdk);

\draw (-4.5, -1.3) rectangle (3.7, 2.0);

\draw (-0.8, 0.5) rectangle (3.5, 1.8);
\node [draw=none,fill=none] (coqlib) at (1.4, 0.8) {Coq standard library};

\draw (-4.3, 0.5) rectangle (-1.3, 1.8);
\node [draw=none,fill=none] (opentheory) at (-2.8, 0.8) {OpenTheory};

\node [manual] (interopdk) at (5.5, 0.5) {\texttt{interop.dk}} edge [dep] (3.7, 0.5);

\end{tikzpicture}
\end{center}
\vspace{-10pt}
\caption{
  Components of the implementation. Solid frames represent source files. Dashed frames represent automatically generated files. Arrows represent dependencies. }
\label{fig:dependencies}
\end{figure}

We chose this example because the interaction between Coq and HOL types is very
limited thanks to polymorphism: there is no need to reason about HOL natural
numbers on the Coq side and no need to reason about lists on the HOL side so the
only interaction takes place at the level of booleans which we wanted to study.
We think it would have been harder for example to translate and link theorems
about natural numbers in HOL and theorems about natural numbers in Coq.
Our implementation is illustrated in Figure \ref{fig:dependencies}.
All components were successfully verified by Dedukti.

\section{Conclusion}

We successfully translated a small Coq development to Dedukti and instanciated
it with the HOL definition of natural numbers. The results have
been validated by Dedukti. Mixing the underlying theories of Coq and HOL raised
interesting questions but did not require a lot of human work: the file
\texttt{hol.dk} is very close to the version included with Holide and the file
\texttt{holtypes.v} is very small. In retrospect, the result looks a lot like
an embedding of HOL in Coq but performed in Dedukti. This is not surprising, as
the theory of HOL is fairly simple compared to Coq and is in fact a subset of
the logic of Coq \cite{barendregt_lambda_1992,geuvers_logics_1993,keller_importing_2010}.

The interoperability layer \texttt{interop.dk} which is
specific to our case study required a lot of work which should be
automated before using this approach on larger scale; our next step on this front
will be to integrate Zenon to solve the proof obligations when they happen to be
in the first-order fragment.
Interoperability raises more issues than mere proof rechecking and our
translators to Dedukti need to be improved. The translations produce
code intended for machines that is not very usable by humans. The
linking of theories together should therefore either be more automated
or benefit from a more readable output. We expect more complex
examples of interoperability to require some form of parametrization
in the translators: when the developer wants the translator to map a
given symbol to a specific Dedukti definition, he should be able to
alter the behaviour of the translator by annotations in some source file,
as done by Keller and Werner \cite{keller_importing_2010} and by Hurd \cite{hurd_opentheory_2011}.

Another limitation of this example of interoperability is the lack of executability.
Even though we have constructed a sorting ``algorithm'' on lists of HOL natural
numbers and we have proved it correct, there is no way to actually execute this
algorithm. Indeed, there is no notion of computation in HOL, so when the sorting
algorithm asks \const{compare} for a comparison between two numbers, it will not
return something which will unblock the computation. Therefore,
$\const{insertion\_sort}\ [4, 1, 3, 2]$ is not \emph{computationally} equal to
$[1, 2, 3, 4]$. However, the result is still \emph{provably} equal to what is
expected: we can show that $\const{insertion\_sort}\ [4, 1, 3, 2]$ is equal to
$[1, 2, 3, 4]$. A constructive and computational presentation of HOL will be necessary before we can
obtain truly executable code. The pure type system presentation of HOL
\cite{barendregt_lambda_1992,geuvers_logics_1993} is a reasonable candidate for
that but the proofs of OpenTheory will need to be adapted. Holide seems like a
good starting point for such a transformation and is the subject of current ongoing work.



\bibliographystyle{eptcs}
\bibliography{interop}

\begin{thebibliography}{10}
\providecommand{\bibitemdeclare}[2]{}
\providecommand{\surnamestart}{}
\providecommand{\surnameend}{}
\providecommand{\urlprefix}{Available at }
\providecommand{\url}[1]{\texttt{#1}}
\providecommand{\href}[2]{\texttt{#2}}
\providecommand{\urlalt}[2]{\href{#1}{#2}}
\providecommand{\doi}[1]{doi:\urlalt{http://dx.doi.org/#1}{#1}}
\providecommand{\bibinfo}[2]{#2}

\bibitemdeclare{unpublished}{assaf_calculus_2014}
\bibitem{assaf_calculus_2014}
\bibinfo{author}{Ali \surnamestart Assaf\surnameend} (\bibinfo{year}{2014}):
  \emph{\bibinfo{title}{{A calculus of constructions with explicit
  subtyping}}}.
\newblock \urlprefix\url{https://hal.inria.fr/hal-01097401}.
\newblock \bibinfo{note}{Accepted in {P}ostproceedings of {Types} 2014}.

\bibitemdeclare{inproceedings}{assaf_conservativity_2015}
\bibitem{assaf_conservativity_2015}
\bibinfo{author}{Ali \surnamestart Assaf\surnameend} (\bibinfo{year}{2015}):
  \emph{\bibinfo{title}{{Conservativity of embeddings in the lambda-{Pi}
  calculus modulo rewriting}}}.
\newblock In \bibinfo{editor}{Thorsten \surnamestart Altenkirch\surnameend},
  editor: {\sl \bibinfo{booktitle}{International Conference on Typed Lambda Calculi and Applications (TLCA)}}, {\sl \bibinfo{series}{LIPIcs}}
  \bibinfo{volume}{38}, \bibinfo{publisher}{Schloss Dagstuhl - Leibniz-Zentrum fuer Informatik}, pp.
  \bibinfo{pages}{31--44}, \doi{10.4230/LIPIcs.TLCA.2015.31}.

\bibitemdeclare{unpublished}{assaf_translating_2014}
\bibitem{assaf_translating_2014}
\bibinfo{author}{Ali \surnamestart Assaf\surnameend} \&
  \bibinfo{author}{Guillaume \surnamestart Burel\surnameend}
  (\bibinfo{year}{2015}): \emph{\bibinfo{title}{{Translating {HOL} to
  {Dedukti}}}}.
\newblock \urlprefix\url{https://hal.inria.fr/hal-01097412}.
\newblock \bibinfo{note}{Accepted in {PxTP} 2015}.

\bibitemdeclare{incollection}{barendregt_lambda_1992}
\bibitem{barendregt_lambda_1992}
\bibinfo{author}{Henk \surnamestart Barendregt\surnameend}
  (\bibinfo{year}{1992}): \emph{\bibinfo{title}{Lambda calculi with types}}.
\newblock In \bibinfo{editor}{Samson \surnamestart Abramsky\surnameend},
  \bibinfo{editor}{Dov~M. \surnamestart Gabbay\surnameend} \&
  \bibinfo{editor}{Thomas S.~E. \surnamestart Maibaum\surnameend}, editors:
  {\sl \bibinfo{booktitle}{Handbook of {Logic} in {Computer} {Science}}},
  \bibinfo{volume}{2}, \bibinfo{publisher}{Oxford University Press}, pp.
  \bibinfo{pages}{117--309}.

\bibitemdeclare{inproceedings}{boespflug_lambda-pi-calculus_2012}
\bibitem{boespflug_lambda-pi-calculus_2012}
\bibinfo{author}{M.~\surnamestart Boespflug\surnameend},
  \bibinfo{author}{Q.~\surnamestart Carbonneaux\surnameend} \&
  \bibinfo{author}{O.~\surnamestart Hermant\surnameend} (\bibinfo{year}{2012}):
  \emph{\bibinfo{title}{The lambda-{Pi}-calculus modulo as a universal proof
  language}}.
\newblock In: {\sl \bibinfo{booktitle}{Proof {Exchange} for {Theorem} {Proving}
  - {Second} {International} {Workshop}, {PxTP} 2012}}, pp.
  \bibinfo{pages}{28--43}.

\bibitemdeclare{inproceedings}{boespflug_coqine_2012}
\bibitem{boespflug_coqine_2012}
\bibinfo{author}{Mathieu \surnamestart Boespflug\surnameend} \&
  \bibinfo{author}{Guillaume \surnamestart Burel\surnameend}
  (\bibinfo{year}{2012}): \emph{\bibinfo{title}{{{CoqInE}: Translating the
  calculus of inductive constructions into the $\lambda\Pi$-calculus modulo}}}.
\newblock In: {\sl \bibinfo{booktitle}{{Proof {Exchange} for {Theorem}
  {Proving} - {Second} {International} {Workshop}, {PxTP} 2012}}},
  p.~\bibinfo{pages}{44}.

\bibitemdeclare{inproceedings}{zenon}
\bibitem{zenon}
\bibinfo{author}{Richard \surnamestart Bonichon\surnameend},
  \bibinfo{author}{David \surnamestart Delahaye\surnameend} \&
  \bibinfo{author}{Damien \surnamestart Doligez\surnameend}
  (\bibinfo{year}{2007}): \emph{\bibinfo{title}{{{Zenon}: An Extensible
  Automated Theorem Prover Producing Checkable Proofs}}}.
\newblock In: {\sl \bibinfo{booktitle}{Logic for Programming Artificial
  Intelligence and Reasoning (LPAR)}}, {\sl \bibinfo{series}{LNCS/LNAI}}
  \bibinfo{volume}{4790}, \bibinfo{publisher}{Springer}, pp.
  \bibinfo{pages}{151--165}, \doi{10.1007/978-3-540-75560-9\_13}.

\bibitemdeclare{inproceedings}{burel_shallow_2013}
\bibitem{burel_shallow_2013}
\bibinfo{author}{Guillaume \surnamestart Burel\surnameend}
  (\bibinfo{year}{2013}): \emph{\bibinfo{title}{A Shallow Embedding of
  Resolution and Superposition Proofs into the $\lambda${$\Pi$}-Calculus
  Modulo}}.
\newblock In \bibinfo{editor}{Jasmin~Christian \surnamestart
  Blanchette\surnameend} \& \bibinfo{editor}{Josef \surnamestart
  Urban\surnameend}, editors: {\sl \bibinfo{booktitle}{{Proof {Exchange} for
  {Theorem} {Proving} - {Third} {International} {Workshop}, {PxTP} 2013}}},
  {\sl \bibinfo{series}{EPiC}}~\bibinfo{volume}{14},
  \bibinfo{publisher}{{EasyChair}}, pp. \bibinfo{pages}{43--57}.

\bibitemdeclare{unpublished}{Zenon-Modulo}
\bibitem{Zenon-Modulo}
\bibinfo{author}{Rapha{\"e}l \surnamestart Cauderlier\surnameend} \&
  \bibinfo{author}{Pierre \surnamestart Halmagrand\surnameend}
  (\bibinfo{year}{2015}): \emph{\bibinfo{title}{Checking {Zenon Modulo} proofs
  in {Dedukti}}}.
\newblock \bibinfo{note}{Accepted in {PxTP} 2015}.

\bibitemdeclare{incollection}{cousineau_embedding_2007}
\bibitem{cousineau_embedding_2007}
\bibinfo{author}{Denis \surnamestart Cousineau\surnameend} \&
  \bibinfo{author}{Gilles \surnamestart Dowek\surnameend}
  (\bibinfo{year}{2007}): \emph{\bibinfo{title}{Embedding Pure Type Systems in
  the Lambda-{P}i-Calculus Modulo}}.
\newblock In \bibinfo{editor}{Simona Ronchi~Della \surnamestart
  Rocca\surnameend}, editor: {\sl \bibinfo{booktitle}{Typed Lambda Calculi and
  Applications, 8th International Conference, {TLCA} 2007, {P}aris, {F}rance,
  June 26-28, 2007, Proceedings}}, {\sl \bibinfo{series}{{LNCS}}}
  \bibinfo{volume}{4583}, \bibinfo{publisher}{Springer}, pp.
  \bibinfo{pages}{102--117},
  \doi{10.1007/978-3-540-73228-0\_9}.

\bibitemdeclare{incollection}{delahaye_zenon_2013}
\bibitem{delahaye_zenon_2013}
\bibinfo{author}{David \surnamestart Delahaye\surnameend},
  \bibinfo{author}{Damien \surnamestart Doligez\surnameend},
  \bibinfo{author}{Fr\'ed\'eric \surnamestart Gilbert\surnameend},
  \bibinfo{author}{Pierre \surnamestart Halmagrand\surnameend} \&
  \bibinfo{author}{Olivier \surnamestart Hermant\surnameend}
  (\bibinfo{year}{2013}): \emph{\bibinfo{title}{Zenon Modulo: When {A}chilles
  Outruns the Tortoise Using Deduction Modulo}}.
\newblock In \bibinfo{editor}{Ken \surnamestart McMillan\surnameend},
  \bibinfo{editor}{Aart \surnamestart Middeldorp\surnameend} \&
  \bibinfo{editor}{Andrei \surnamestart Voronkov\surnameend}, editors: {\sl
  \bibinfo{booktitle}{{LPAR}}}, {\sl \bibinfo{series}{{LNCS}}}
  \bibinfo{volume}{8312}, \bibinfo{publisher}{Springer Berlin Heidelberg}, pp.
  \bibinfo{pages}{274--290},
  \doi{10.1007/978-3-642-45221-5\_20}.

\bibitemdeclare{techreport}{dowek_models_2014}
\bibitem{dowek_models_2014}
\bibinfo{author}{Gilles \surnamestart Dowek\surnameend} (\bibinfo{year}{2014}):
  \emph{\bibinfo{title}{Models and termination of proof-reduction in the
  lambda-{Pi}-calculus modulo theory}}.
\newblock \bibinfo{type}{Technical report}, \bibinfo{institution}{Inria},
  \bibinfo{address}{Paris}.
\newblock
  \urlprefix\url{https://who.rocq.inria.fr/Gilles.Dowek/Publi/superpi.pdf}.

\bibitemdeclare{phdthesis}{geuvers_logics_1993}
\bibitem{geuvers_logics_1993}
\bibinfo{author}{Herman \surnamestart Geuvers\surnameend}
  (\bibinfo{year}{1993}): \emph{\bibinfo{title}{Logics and type systems}}.
\newblock \bibinfo{type}{{PhD} thesis}, \bibinfo{school}{University of
  Nijmegen}.

\bibitemdeclare{article}{harper_framework_1993}
\bibitem{harper_framework_1993}
\bibinfo{author}{Robert \surnamestart Harper\surnameend},
  \bibinfo{author}{Furio \surnamestart Honsell\surnameend} \&
  \bibinfo{author}{Gordon \surnamestart Plotkin\surnameend}
  (\bibinfo{year}{1993}): \emph{\bibinfo{title}{A framework for defining
  logics}}.
\newblock {\sl \bibinfo{journal}{Journal of the ACM}}
  \bibinfo{volume}{40}(\bibinfo{number}{1}), pp. \bibinfo{pages}{143--184},
  \doi{10.1145/138027.138060}.

\bibitemdeclare{incollection}{harrison_hol_2009}
\bibitem{harrison_hol_2009}
\bibinfo{author}{John \surnamestart Harrison\surnameend}
  (\bibinfo{year}{2009}): \emph{\bibinfo{title}{{HOL} {Light}: An Overview}}.
\newblock In \bibinfo{editor}{Stefan \surnamestart Berghofer\surnameend},
  \bibinfo{editor}{Tobias \surnamestart Nipkow\surnameend},
  \bibinfo{editor}{Christian \surnamestart Urban\surnameend} \&
  \bibinfo{editor}{Makarius \surnamestart Wenzel\surnameend}, editors: {\sl
  \bibinfo{booktitle}{Theorem Proving in Higher Order Logics}}, {\sl
  \bibinfo{series}{{LNCS}}} \bibinfo{volume}{5674},
  \bibinfo{publisher}{Springer Berlin Heidelberg}, pp. \bibinfo{pages}{60--66},
  \doi{10.1007/978-3-642-03359-9\_4}.

\bibitemdeclare{article}{horozalrabe2011}
\bibitem{horozalrabe2011}
\bibinfo{author}{Fulya \surnamestart Horozal\surnameend} \&
  \bibinfo{author}{Florian \surnamestart Rabe\surnameend}
  (\bibinfo{year}{2011}): \emph{\bibinfo{title}{{Representing model theory in a
  type-theoretical logical framework}}}.
\newblock {\sl \bibinfo{journal}{Theoretical Computer Science}}
  \bibinfo{volume}{412}, pp. \bibinfo{pages}{4919--4945},
  \doi{10.1016/j.tcs.2011.03.022}.

\bibitemdeclare{incollection}{hurd_opentheory_2011}
\bibitem{hurd_opentheory_2011}
\bibinfo{author}{Joe \surnamestart Hurd\surnameend} (\bibinfo{year}{2011}):
  \emph{\bibinfo{title}{{The {OpenTheory} Standard Theory Library}}}.
\newblock In \bibinfo{editor}{Mihaela \surnamestart Bobaru\surnameend},
  \bibinfo{editor}{Klaus \surnamestart Havelund\surnameend},
  \bibinfo{editor}{Gerard~J. \surnamestart Holzmann\surnameend} \&
  \bibinfo{editor}{Rajeev \surnamestart Joshi\surnameend}, editors: {\sl
  \bibinfo{booktitle}{{{NFM}}}}, {\sl \bibinfo{series}{{{LNCS}}}}
  \bibinfo{volume}{6617}, \bibinfo{publisher}{Springer}, pp.
  \bibinfo{pages}{177--191},
  \doi{10.1007/978-3-642-20398-5\_14}.

\bibitemdeclare{incollection}{kaliszyk_scalable_2013}
\bibitem{kaliszyk_scalable_2013}
\bibinfo{author}{Cezary \surnamestart Kaliszyk\surnameend} \&
  \bibinfo{author}{Alexander \surnamestart Krauss\surnameend}
  (\bibinfo{year}{2013}): \emph{\bibinfo{title}{Scalable {LCF}-style proof
  translation}}.
\newblock In \bibinfo{editor}{Sandrine \surnamestart Blazy\surnameend},
  \bibinfo{editor}{Christine \surnamestart Paulin-Mohring\surnameend} \&
  \bibinfo{editor}{David \surnamestart Pichardie\surnameend}, editors: {\sl
  \bibinfo{booktitle}{Interactive {Theorem} {Proving}}}, {\sl
  \bibinfo{series}{{LNCS}}} \bibinfo{volume}{7998},
  \bibinfo{publisher}{Springer Berlin Heidelberg}, pp. \bibinfo{pages}{51--66},
  \doi{10.1007/978-3-642-39634-2\_7}.

\bibitemdeclare{incollection}{keller_importing_2010}
\bibitem{keller_importing_2010}
\bibinfo{author}{Chantal \surnamestart Keller\surnameend} \&
  \bibinfo{author}{Benjamin \surnamestart Werner\surnameend}
  (\bibinfo{year}{2010}): \emph{\bibinfo{title}{Importing {HOL} {Light} into
  {Coq}}}.
\newblock In \bibinfo{editor}{Matt \surnamestart Kaufmann\surnameend} \&
  \bibinfo{editor}{Lawrence~C. \surnamestart Paulson\surnameend}, editors: {\sl
  \bibinfo{booktitle}{{ITP}}}, {\sl \bibinfo{series}{{LNCS}}}
  \bibinfo{volume}{6172}, \bibinfo{publisher}{Springer Berlin Heidelberg}, pp.
  \bibinfo{pages}{307--322},
  \doi{10.1007/978-3-642-14052-5\_22}.

\bibitemdeclare{incollection}{naumov_hol/nuprl_2001}
\bibitem{naumov_hol/nuprl_2001}
\bibinfo{author}{Pavel \surnamestart Naumov\surnameend},
  \bibinfo{author}{Mark-Oliver \surnamestart Stehr\surnameend} \&
  \bibinfo{author}{Jos\'e \surnamestart Meseguer\surnameend}
  (\bibinfo{year}{2001}): \emph{\bibinfo{title}{The {HOL}/{NuPRL} proof
  translator}}.
\newblock In \bibinfo{editor}{Richard~J. \surnamestart Boulton\surnameend} \&
  \bibinfo{editor}{Paul~B. \surnamestart Jackson\surnameend}, editors: {\sl
  \bibinfo{booktitle}{Theorem {Proving} in {Higher} {Order} {Logics}}}, {\sl
  \bibinfo{series}{{LNCS}}} \bibinfo{volume}{2152},
  \bibinfo{publisher}{Springer Berlin Heidelberg}, pp.
  \bibinfo{pages}{329--345}, \doi{10.1007/3-540-44755-5\_23}.

\bibitemdeclare{incollection}{obua_importing_2006}
\bibitem{obua_importing_2006}
\bibinfo{author}{Steven \surnamestart Obua\surnameend} \&
  \bibinfo{author}{Sebastian \surnamestart Skalberg\surnameend}
  (\bibinfo{year}{2006}): \emph{\bibinfo{title}{Importing {HOL} into
  {Isabelle}/{HOL}}}.
\newblock In \bibinfo{editor}{Ulrich \surnamestart Furbach\surnameend} \&
  \bibinfo{editor}{Natarajan \surnamestart Shankar\surnameend}, editors: {\sl
  \bibinfo{booktitle}{Automated {Reasoning}}}, {\sl \bibinfo{series}{{LNCS}}}
  \bibinfo{volume}{4130}, \bibinfo{publisher}{Springer Berlin Heidelberg}, pp.
  \bibinfo{pages}{298--302}, \doi{10.1007/11814771\_27}.

\bibitemdeclare{incollection}{pfenning_system_1999}
\bibitem{pfenning_system_1999}
\bibinfo{author}{Frank \surnamestart Pfenning\surnameend} \&
  \bibinfo{author}{Carsten \surnamestart Sch{\"u}rmann\surnameend}
  (\bibinfo{year}{1999}): \emph{\bibinfo{title}{System Description: {Twelf}
  {\textemdash} A Meta-Logical Framework for Deductive Systems}}.
\newblock In: {\sl \bibinfo{booktitle}{Automated Deduction {\textemdash}
  {CADE}-16}}, {\sl \bibinfo{series}{{LNCS}}} \bibinfo{volume}{1632},
  \bibinfo{publisher}{Springer Berlin Heidelberg}, pp.
  \bibinfo{pages}{202--206}, \doi{10.1007/3-540-48660-7\_14}.

\bibitemdeclare{inproceedings}{saillard_dedukti_2013}
\bibitem{saillard_dedukti_2013}
\bibinfo{author}{Ronan \surnamestart Saillard\surnameend}
  (\bibinfo{year}{2013}): \emph{\bibinfo{title}{Dedukti: a universal proof
  checker}}.
\newblock In: {\sl \bibinfo{booktitle}{Foundation of {Mathematics} for
  {Computer}-{Aided} {Formalization} {Workshop}}}, \bibinfo{address}{Padova}.
\newblock \urlprefix\url{https://hal.inria.fr/hal-00833992}.

\bibitemdeclare{inproceedings}{saillard_towards_2013}
\bibitem{saillard_towards_2013}
\bibinfo{author}{Ronan \surnamestart Saillard\surnameend}
  (\bibinfo{year}{2013}): \emph{\bibinfo{title}{{Towards explicit rewrite rules
  in the $\lambda\Pi$-calculus modulo}}}.
\newblock In: {\sl \bibinfo{booktitle}{{IWIL} - 10th International Workshop on
  the Implementation of Logics}}.
\newblock \urlprefix\url{https://hal.inria.fr/hal-00921340}.

\bibitemdeclare{incollection}{schurmann_executable_2006}
\bibitem{schurmann_executable_2006}
\bibinfo{author}{Carsten \surnamestart Sch{\"u}rmann\surnameend} \&
  \bibinfo{author}{Mark-Oliver \surnamestart Stehr\surnameend}
  (\bibinfo{year}{2006}): \emph{\bibinfo{title}{An Executable Formalization of
  the {HOL}/{Nuprl} Connection in the Metalogical Framework Twelf}}.
\newblock In \bibinfo{editor}{Miki \surnamestart Hermann\surnameend} \&
  \bibinfo{editor}{Andrei \surnamestart Voronkov\surnameend}, editors: {\sl
  \bibinfo{booktitle}{Logic for Programming, Artificial Intelligence, and
  {Reasoning}}}, {\sl \bibinfo{series}{{LNCS}}} \bibinfo{volume}{4246},
  \bibinfo{publisher}{Springer Berlin Heidelberg}, pp.
  \bibinfo{pages}{150--166}, \doi{10.1007/11916277\_11}.

\bibitemdeclare{incollection}{TPTPderivation}
\bibitem{TPTPderivation}
\bibinfo{author}{Geoff \surnamestart Sutcliffe\surnameend},
  \bibinfo{author}{Stephan \surnamestart Schulz\surnameend},
  \bibinfo{author}{Koen \surnamestart Claessen\surnameend} \&
  \bibinfo{author}{Allen \surnamestart Van~Gelder\surnameend}
  (\bibinfo{year}{2006}): \emph{\bibinfo{title}{Using the TPTP Language for
  Writing Derivations and Finite Interpretations}}.
\newblock In \bibinfo{editor}{Ulrich \surnamestart Furbach\surnameend} \&
  \bibinfo{editor}{Natarajan \surnamestart Shankar\surnameend}, editors: {\sl
  \bibinfo{booktitle}{Automated Reasoning}}, {\sl \bibinfo{series}{{LNCS}}}
  \bibinfo{volume}{4130}, \bibinfo{publisher}{Springer Berlin Heidelberg}, pp.
  \bibinfo{pages}{67--81}, \doi{10.1007/11814771\_7}.

\bibitemdeclare{manual}{coq_development_team_coq_2012}
\bibitem{coq_development_team_coq_2012}
\bibinfo{author}{The~{Coq} \surnamestart development team\surnameend}
  (\bibinfo{year}{2012}): \emph{\bibinfo{title}{The {Coq} Reference Manual,
  version 8.4}}.
\newblock \urlprefix\url{http://coq.inria.fr/doc}.

\end{thebibliography}
\end{document}